\documentstyle[12pt]{article}

\begin{document}

\begin{flushright}
{\bf hep-th/9505118}\\
{\bf US--FT/6--95}\\
{\bf May 1995}
\end{flushright}

\vspace{1cm}

\begin{center}
 {\bf {\huge Integrable Systems and W-algebras }}
 \vspace{0.4cm}

 { C. R. Fern\'andez-Pousa, M. V. Gallas, J. L. Miramontes,\\
  and J. S\'anchez Guill\'en}

 \vspace{0.3cm}
 {\it \small{ Departamento de F\'{\i}sica de Part\'{\i}culas,
 Universidade de Santiago de Compostela,\\
 15706--Santiago de Compostela, Spain}}

 \vspace{0.3cm}
 {\it \small{ Talk given by J. S. G.\footnote{JOAQUIN@GAES.USC.ES}
 at the VIIIth J. A. Swieca Summer School,\\
 Rio de Janeiro, Brazil, February 1995}}
\end{center}


\begin{abstract}
 {\small The basic concepts underlying our analysis of {\it W-algebras}
 as extended symmetries of integrable systems are summarized. The
 construction starts from the  second hamiltonian structure of
 ``Generalized Drinfel'd-Sokolov'' hierarchies,  and its
 correspondence with the $A_1$-embeddings is established, providing a
 rather simple and general scheme.}
\end{abstract}

We summarize the building blocks of our  results$^1$ about the
construction and classification of {\it W-algebras}$^2$ starting
from the ``Generalized  Drinfel'd-Sokolov'' (G.D-S) integrable non-linear
hierarchies$^3$, and  their relation to the method of hamiltonian
reduction and $A_{1}$-embeddings$^4$.

Classical {\it W-algebras} are non-linear chiral extensions of the
conformal Virasoro algebra generated by primary fields $w_i(z)$:
\begin{eqnarray}
 \{ w_{i}(z_{1}), w_{j}(z_{2})\} & = & \sum_{k}
    P^{k}_{ij}(w_{1},....w_{N}) \,\,\delta^{(k)}(z_{1}-z_{2}) \> ,
\end{eqnarray}
where $\{,\}$ is a Poisson bracket, and $P^{k}_{ij}$ are differential
polynomials constrained by the Jacobi identity. The energy-momentum
tensor is assigned to $i=1$
\begin{eqnarray}
 \{ T(z_{1}), w_{i}(z_{2})\}  = & - \, \Delta_{i}\, w_{i}(z_{1})\,
 \delta'(z_{1}-z_{2}) \,- (\Delta_{i}-1) \, w'_{i}(z_{1})\,
 \delta(z_{1}-z_{2})
\end{eqnarray}
and  $\Delta_{i}$ is the conformal weight of $w_{i}(z)$.

{\it W-algebras} appear in many different applications of two-dimensional
classical and quantum field theory, such as String Theory, Gravity,
Critical Phenomena, or Solid State,  and offer good prospects for
combining external and internal symmetries in  theories of  higher spins.
{\it W-algebras} generally describe symmetries of two-dimensional
integrable systems. In particular, they correspond to the ``second''
hamiltonian structure of G.D-S hierarchies,  which can be related in a
simple way to the general construction based on $A_{1}$-embeddings.

We remind that the second hamiltonian structure of the K.d.V. hierarchies
associated to the affine Kac-Moody algebras  $A^{(1)}_{n}$ $^5$ lead to
the well known  $W_{n+1}$ algebras, the first of  which is, of course, the
Virasoro algebra:
\begin{eqnarray}
 \{u(x,t),u(y,t)\}_{2} = \frac{1}{2} \, \delta'''(x-y) - 2u(x,t)\,
 \delta'(x-y) -  u'(x,t)\, \delta(x-y) \>,
\end{eqnarray}
which corresponds to $n = 1$ and, hence, to the original K.d.V. equation
\begin{eqnarray}
 \partial_{t} u(x,t) \,& =& -\,\, \frac{1}{4}\, u'''(x,t)\, + \,
           \frac{3}{2}\, u(x,t) u'(x,t)\\
                      &=&\{\int dx \frac{u^{2}(x,t)}{4},u(y,t) \}_{2} \ .
\nonumber
\end{eqnarray}
In general,  the second  Poisson-bracket of the G.D-S hierarchies provides
different {\it W-algebras} associated to distinguished elements of every
affine algebra, and the conformal  structure arises when the well known
scale invariance of the equations is made local.  It is convenient to
explain one of the basic ingredients of the construction: the
Heisenberg  subalgebras (H.S.A) of Kac-Moody algebras, which, in practice,
will be  considered just as loop  algebras,
\begin{eqnarray}
 \hat{g}^{(1)}\, =\, {\bf C}[z,z^{-1}] \bigotimes \, g \,  \bigoplus
                  {\bf C} \, d \>,
\end{eqnarray}
since the central extension ${\bf C} c $ can and will be ignored. $g$ is
the underlying  finite Lie  algebra, and  ${\bf C}[z,z^{-1}] $ are the
Laurent polynomials in the affine parameter $z$. The standard
derivation  $d= z\frac{d}{dz}$  induces a gradation of $\hat{g}^{(1)}$,
which generally means  a decomposition:
\begin{eqnarray}
 \hat{g}^{(1)}\,=\, \bigoplus_{j} \, \hat{g}^{(1)}_{j} \> ,  & \, \,
[\hat{g}^{(1)}_{j},\hat{g}^{(1)}_{k}] \subset \,
          \hat{g}^{(1)}_{j+k}.
\end{eqnarray}
Notice that  $[d,\hat{g}^{(1)}_{j}]=j \, \hat{g}^{(1)}_{j}$
if  $\hat{g}^{(1)}_{j} =z^{j}
\bigotimes g  $ and, so, this distinguished gradation is called {\it
homogeneous}. Also if
$H$ is a Cartan subalgebra, $z^{j} \bigotimes H$ is a  maximal commuting
subalgebra: the {\it homogeneous} example of the {\bf Heisenberg
subalgebras}. General H.S.A's correspond to all the  possible
generalizations of such structure, which are in one to one  relation
with the conjugacy classes of the Weyl group  of $g$.

For each conjugacy class $[w]$ there is an associated  set of $r+1$
(r is the rank of g) coprime  non-negative integers called
$\vec{s}_{w}$,  which will specify the grades under the new  derivation
$d_{s_{w}} \,= \, N_{s} d + H_{s_{w}}\>,$ with $N_{s_{w}} =
\sum k_{i} s^{i}_{w} $  ($k_{i}$ being the Kac labels, i.e., the
components of the right null eigenvector of the extended Cartan matrix),
and $\alpha_{i}(H_{s_{w}})= s^{i}_{w}$ . The way to prove that this
actually  generates all the  inequivalent Heisenberg  subalgebras,
which  will be labelled as ${\cal H}[w]$, is to  express an  inner
automorphism of $g$ associated to $[w]$ as an   imaginary-phase
operator specified  by the adjoint action of $H_{s_{w}}$, which can be
expressed as
$\sum_{i} \frac{2}{{\alpha_i}^{2}} s_{w}^{i} w_{i} H$, with $w_{i}$
being  the fundamental weights$^1$. The {\it homogeneous} case
corresponds to the identity $w= 1$ and $\vec{s}_{w}=(1,0,...,0)$, while
the product of all the Weyl reflections, i.e., the Coxeter element $w_c$,
corresponds to the  {\it principal} gradation. For $sl(2,{\bf C})$ these
are the only cases, but, for instance, in  $A_{2}$ there is also the
$\vec{s}_{w}= (2,1,1)$, which yields the Heisenberg subalgebra generated
by:
\begin{eqnarray}
 b_{4m}= z^{m} \left(\begin{array}{ccc}
        1 &  & \\
          &-2& \\
          &  &1
      \end{array}
\right ) \> ;\quad &  \,  b_{2+4m} = z^{m}\left(\begin{array}{ccc}
        & &1 \\
        &0& \\
        z& &
      \end{array} \right )
         \> ;\quad  &  m \in {\bf Z}.
\end{eqnarray}
The name of these  algebras comes from their central extension ignored
here, where the different elements $X_{p}$ , $[d_{s_{w}}, X_{p}]\,=\, p
X_{p}\,$, would satisfy:
\begin{eqnarray}
 [X_{p},X_{q}] \,=\,p \, \delta_{p+q,0}\,\> c \> .
\end{eqnarray}
Notice that  they are maximally commuting for $c = 0$, and the
suggestive similarity of $d_{s_{w}}$ with the momentum. In fact the former
relations  can be seen as the (2-dimensional) Poincar\'e algebra.

We are now in position to summarize the G.D-S construction of non-linear
differential hierarchies. A Lax operator is then defined for every
constant element of positive $s_{w}$-grade, $\Lambda^{i} \in
{\cal H}[w]$,
\begin{eqnarray}
 L\, = \, \partial_{x} + \Lambda^{i}(z) + q(x,z) \>, && q(x,z) \in
 Q^{<i}_{\geq 0} \equiv \hat{g}_{\geq 0}(s) \bigcap
 \hat{g}_{<i}(s_{w})
\end{eqnarray}
above, $s$ is some other gradation of $\hat{g}$ such that
$s\leq  s_{w}$. The partial ordering implies$^3$ that $ \hat{g}_{j(s_w)}
\subset {\hat g}_{j(s)}$,  for $j \geq 0 $, $j \leq 0$ and  $j = 0$.

$Q$, the  phase space the of hierarchy, is invariant under the {\it gauge
transformations} generated by the adjoint action of  $S =
{\hat g}^{<0}_{0}$. Moreover, as long as $Ker(ad\Lambda)= {\cal H}[w]$
\footnote{A condition which  can be relaxed for non-regular elements of
${\cal H}[w]$.}, one can ``diagonalize'' $L$  in ${\cal H}[w]$ by
means of an element
$V \in Im(ad \Lambda) \bigcap {\hat g}^{< 0}$ :
\begin{eqnarray}
 {\cal L} \,= e^{ad V}\, (L)\, = \, \partial_{x} + \Lambda^{i}(z) +
\sum_{j<i}
         H_{j}(x)
\end{eqnarray}

Since  $[M_{+},L]= - [M_{-},L] \in
C^{\infty }({\bf R}, Q)$ for  $M$ commuting with L, one can define
the flows:
\begin{eqnarray}
 \frac{\partial L}{ \partial t_{b_{j}}} \, =  [(exp(-ad V)(b_{j}))^{\geq
                0_{s_{w}}}, L] =\,- \,
               [(exp(-ad V)(b_{j}))^{< 0_{s_{w}}},L] \>,
\end{eqnarray}
$b_{j} \in {\cal H}[w] $, or through  similar equations involving  of the
lower gradation $s$.

Provided that $Ker (ad \Lambda) \bigcap S =\emptyset $ (injective gauge
transformations), the non-linear differential equations for  the {\it
(canonical)} gauge invariant functionals $q^{can}$ will be polynomial,
as required by locality in the field theory applications, then,
$Q^{can}$  will be just a  complementary space of  $[\Lambda , S]$.
For  regular elements, $Ker (ad \Lambda) \bigcap S =\emptyset $
is always satisfied (our$^1$ Lemma 1).

By construction, the $H_{j}$ for $j < 0$ are the conserved densities of
the hierarchy (for $i>j \geq 0$ they are just centers), the hierarchy
can be written in zero curvature form, and more important, it is
bihamiltonian and the second Poisson bracket  exhibits conformal
invariance.

It is then natural to investigate whether such structures are related to
the W-algebras obtained by hamiltonian reduction of the affine Kac-Moody
current algebras. Moreover,  the latter are classified by the
$A_{1}$-embeddings, which are subalgebras $\{ J_{\pm}, J_{0}\}$ of $g$
such that, under their ad-action, $g$ decomposes into (spin) irreducible
representations of that $sl(2,{\bf C})$.  Their $J_{+}$  are related to
the reduced currents $J^{red}$ by addition of  minimal weights of the
spin  decomposition. In $A_{n}$ the embeddings are in correspondence with
the Weyl group, and they are also labeled by partitions of $n$.
 So, one expects them to be related to the hierarchies arising from
Heisenberg subalgebras. The precise way in which this happens  is one
of  the results of our paper$^1$. The first step is to derive a
generator of the conformal transformations that generalize the local
scale  invariance of the hierarchy, $q^{k} \rightarrow \lambda^{k/i-1}
q^{k}$ when $x \rightarrow \lambda x $ and $t _{b^{j}} \rightarrow
\lambda^{j/i} t_{ b^{j}}$. Such an energy momentum tensor is
easily derived upon restriction to the functionals  of zero lower grade
$q_{0}(x)$ since, for them, the second  Poisson bracket simplifies to  the
Kirillov bracket
\begin{eqnarray}
 \{\varphi ,\psi\}_{2} &= & \left( [(d_{q} \varphi)_{0},(d_{q}
 \psi)_{0}], (\Lambda^{i} + q(x))_{0} \right) - \left((d_{q}
 \varphi)_{0},\partial_{x} (d_{q} \psi)_{0} \right ) \>,
\end{eqnarray}
where $(,)$ is  the generalization of the Killing form $<,>$ to the
affine algebra
\begin{eqnarray}
 (A,B)\, = \, \sum _{k\in {\bf Z}}  \int dx  <A_{k}(x),B_{-k}(x)>
\end{eqnarray}
and $d_{q} $ is the Frechet derivative.

Restricting further the construction to the case when
$Ker(ad \Lambda_{0}) \bigcap S\,= \emptyset $, one shows directly  that
the second bracket is a W-algebra. A previous step is to  show, by a
counting argument,  that the  choice of $q^{can}$ is such that the
restriction of the positive grade components $(q^{can>0} (x))_{0} = 0$,
and  that those of  $q^{can \, \leq 0}(x)$ depend only on $q_{0}(x)$. The
embedding corresponding to the W-algebra associated to $\{ \hat{g},
\Lambda^{i>0}
\in  {\cal H}[w], s_{w}, s \}$ is specified by $J_{+}=(\Lambda)_{0}$.
The   reverse of this result$^1$, Theorem 2,  is the answer to the
question of which  $J_{+}$ are obtained in this way from the
corresponding Heisenberg subalgebra. The solution turns out quite
restrictive, and it is given$^1$ by Theorem 3:
for $A_{n}$, the {\it W-algebras} that can be constructed from Heisenberg
subalgebras with the G.D-S formalism  are just those corresponding to
the embeddings labelled by  partitions of the form $n+1 = k(m)+q(1)$ or
$n+1 = k(m+1) + k(m) + q(1)$, where $k$ and $q$  indicate the times the
integer in the bracket is repeated.



\vskip 0.5cm

{\Large{\bf References.}}\newline
{\small
1. C. R. Fern\'andez-Pousa,  M. V. Gallas, J. L. Miramontes and
J. S\'anchez Guill\'en US-FT/13-94 preprint, hep-th/9409016, (to be
published in Ann. Phys.)
\newline
2. A.B. Zamolodchikov, Theor. Math. Phys. {\bf 65} (1985), 347.
\newline
3. T. J. Hollowood and J. L. Miramontes, Commun. Math. Phys.
{\bf 157} (1993) 99.
\newline
4. L. Feh\'er, L. O'Raifeartaigh, P. Ruelle, I. Tsutsui, and A. Wipf,
Phys. Rep. {\bf 222} (1992) 1.
\newline
5. V. G. Drinfel'd and V. V. Sokolov, J. Sov. Math. {\bf 30} (1985) 1975:
Soviet. Math. Dokl. {\bf 23} (1981) 457.
\newline
6. V. G. Kac, Infinite Dimensional Lie Algebras ($3^rd$ ed.), Cambridge
University Press, Cambridge (1990), chap. 8.
\newline
}
\end{document}